\documentclass[12pt]{article}
\usepackage{epsfig}
\usepackage{amsfonts}
\begin{document}
\begin{titlepage}
\begin{center}

{\Large Superstatistics in hydrodynamic turbulence}

\vspace{2.cm}

{\bf Christian Beck}

\vspace{0.5cm}

School of Mathematical Sciences, Queen Mary, University of London,
Mile End Road, London E1 4NS.

\end{center}

\vspace{4cm}

\abstract{Superstatistics is a `statistics of a statistics'
relevant for driven nonequilibrium systems with fluctuating
intensive parameters. It contains Tsallis statistics as a special
case. We show that probability density functions of velocity
differences and accelerations measured in Eulerian and Lagrangian
turbulence experiments are well reproduced by simple
superstatistics models. We compare fits obtained
for log-normal superstatistics and $\chi^2$-superstatistics
($=$ Tsallis statistics). 
}

\vspace{1.3cm}

\end{titlepage}

\section{Introduction}

There is currently considerable interest in more
general versions of statistical mechanics, known under the name
nonextensive statistical mechanics \cite{tsa1,tsa2,abe}. In the mean time, it
has become clear that Tsallis' original approach
\cite{tsa1}
can generalized
in various ways, and that these techniques are often
relevant for the effective description of nonequilibrium systems
with strong fluctuations of an intensive parameter, where ordinary
statistical mechanics has little to say
\cite{wilk,prl,eddie}. A particular class of
more general statistics relevant for nonequilibrium systems,
containing Tsallis statistics as a special case, has been termed
'superstatistics' \cite{eddie,souza,eddie2}. A superstatistics arises out
 of the superposition of two statistics,
namely one described by ordinary Boltzmann factors $e^{-\beta E}$
and another one given by the probability distribution of $\beta$.
This means the inverse temperature parameter $\beta$ is assumed
not to be constant but to be fluctuating on a relatively large
time scale or spatial scale. Naturally, this kind of approach is
physically relevant for driven nonequilibrium systems with
fluctuations, rather than for equilibrium systems.

Depending on the probability distribution of $\beta$, there are
infinitely many superstatistics. It has been shown that Tsallis
statistics is a particular superstatistics obtained under the
assumption that $\beta$ is $\chi^2$-distributed
\cite{prl}. Various other
examples of superstatistics have been studied \cite{eddie}, among them
superstatistics of log-normal type. A main result of \cite{eddie} was
that for small $E$ all superstatistics behave in a universal way,
i.e.\ they generate probability distributions close to Tsallis
distributions. But for large $E$ the various superstatistics can
have quite different properties.

In this paper we work out the application of this very new concept
of statistical mechanics to fully developed turbulence. 
By comparison with various data from Eulerian and Lagrangian
turbulence experiments, as well as data from direct numerical
simulations (DNS) of the Navier Stokes equation, we will provide
evidence that superstatistics of log-normal type quite well
describes measured probability densities in hydrodynamic
turbulence. This superstatistics can be dynamically realized by
considering a class of stochastic differential equations
previously introduced in \cite{prl}, but now with a log-normal rather
than $\chi^2$-distribution of the damping parameter. In general, if the velocity
difference is not too large, most superstatistics models yield
probability densities that are similar to Tsallis statistics.
Significant differences only occur for the tails of the
distribution, i.e.\ for very rare events. For the extreme tails,
superstatistics based on log-normal distributions seems to provide
better fits than superstatistics based on $\chi^2$-distributions
(i.e.\ ordinary Tsallis statistics). On the other hand, if the
velocity difference is not too large (say less than about 30
standard deviations), then Tsallis statistics is quite a good
approximation, with the advantage that an explicit formula for the
densities can be given. 
For
`early' work emphasizing the relevance of Tsallis statistics in 3d-turbulence,
see e.g.\ \cite{ramos, ari, hydro}.

We will analyse data sets from three different
experiments/simulations. The experimental measurements were done
by Swinney et al. (Eulerian turbulence) and Bodenschatz et al.
(Lagrangian turbulence). The direct numerical simulation (DNS)
data were obtained by Gotoh et al. We are very grateful to all
three groups for providing us with their data.

\section{Superstatistics and its dynamical realizations}

\subsection{The basic concept}

Let us give a short introduction to the `superstatistics' concept
\cite{eddie}.
The idea is actually applicable to many systems,
not only to turbulent systems. Consider a driven nonequilibrium
systems with spatio-temporal fluctuations of an intensive
parameter $\beta$. This can e.g.\ be the inverse temperature, or a
chemical potential, or a function of the fluctuating energy
dissipation in the flow (for the turbulence application). Locally,
i.e.\ in cells where $\beta$ is approximately constant, the system
is described by ordinary statistical mechanics, i.e.\ ordinary
Boltzmann factors $e^{-\beta E}$, where $E$ is an effective energy
in each cell. To describe the system in the long-term run, one has
to do a spatio-temporal average over the fluctuating $\beta$. One obtains
a superposition of two statistics (that of
$\beta$ and that of $e^{-\beta E}$), hence the name `superstatistics'.
One
may define an effective Boltzmann factor $B(E)$ given by
\begin{equation}
B(E) =\int_0^\infty f(\beta) e^{-\beta E},
\end{equation}
where $f(\beta)$ is the probability distribution of $\beta$. For
type-A superstatistics, one normalizes this effective Boltzmann
factor, obtaining the stationary probability distribution
\begin{equation}
p(E)=\frac{1}{Z}B(E),
\end{equation}
where
\begin{equation}
Z=\int_0^\infty B(E)dE.
\end{equation}
For type-B superstatistics, one includes the
$\beta$-dependent normalization constant into the averaging
process, obtaining
\begin{equation}
p(E)=\int_0^\infty f(\beta) \frac{1}{Z(\beta)}e^{-\beta E}d\beta ,
\end{equation}
where $Z(\beta)$ is the normalization constant of $e^{-\beta E}$
for a given $\beta$. Both approaches can be easily mapped into
each other, by defining a new probability density
$\tilde{f}(\beta)\sim f(\beta)/Z(\beta)$. It is obvious that
Type-B superstatistics with $f$ is equivalent to type-A
superstatistics with $\tilde f$.

A simple dynamical realization of a superstatistics can be
constructed by considering stochastic differential equations with
spatio-temporally fluctuating parameters \cite{prl}. Consider the Langevin
equation
\begin{equation}
\dot{u}= \gamma F(u) +\sigma L(t), \label{1}
\end{equation}
where $L(t)$ is Gaussian white noise, $\gamma >0$ is a friction
constant, $\sigma$ describes the strength of the noise, and
$F(u)=-\frac{\partial}{\partial u} V(u)$ is a drift force. If
$\gamma$ and $\sigma$ are constant then the stationary probability
density of $u$ is proportional to $e^{-\beta V(u)}$, where
$\beta:=\frac{\gamma}{\sigma^2}$ can be identified with the
inverse temperature of ordinary statistical mechanics. Most
generally, however, we may let the parameters $\gamma$ and
$\sigma$ fluctuate so that
$\beta=\frac{\gamma}{\sigma^2}$ has probability density
$f(\beta)$. These fluctuations are assumed to be on a long time
scale so that the system can temporarily reach local equilibrium.
In this case one obtains for the conditional probability
$p(u|\beta)$ (i.e. the probability of $u$ given some value of
$\beta$)
\begin{equation}
p(u|\beta)=\frac{1}{Z(\beta)}\exp \left\{ -\beta V(u)\right\},
\end{equation}
for the joint probability $p(u,\beta)$ (i.e. the probability
to observe both a certain value of $u$ and a certain value of $\beta$)
\begin{equation}
p(u,\beta)=p(u|\beta)f(\beta)
\end{equation}
and for the marginal probability $p(u)$ (i.e.\ the probability
to observe a certain value of $u$ no matter what $\beta$ is)
\begin{equation}
p(u)=\int_0^\infty  p(u|\beta)f(\beta)d\beta . \label{9}
\end{equation}
This marginal distribution is the generalized canonical
distribution of the superstatistics considered. The above
formulation corresponds to type-B superstatistics.

\subsection{Application to turbulent systems}

In the turbulence application, the mathematics is the same as
outlined above, just the physical meaning of the variables
$u,\beta, E$ etc.\ is slightly different from
that of an ordinary Brownian particle. First of all, $u$ stands
for a local velocity {\em difference} in the turbulent flow. On a very small
time scale, this velocity difference is essentially the
acceleration. It is really the velocity difference, not the
velocity itself that we want to understand. Velocity differences
in turbulence
have been the subject of intensive investigations since the early
work of Kolmogorov. The velocity itself is known to be
approximately Gaussian, so we don't need any sophisticated model
to understand this.

The basic idea is that turbulent velocity
differences locally relax with a certain damping constant $\gamma$
and are at the same time driven by rapidly fluctuating chaotic
force differences. As a local momentum balance, we thus end up
with eq.~(\ref{1}), where we model the chaotic force differences
by Gaussian white noise. As has been shown in \cite{hilgers1,hilgers2}, 
this approximation by Gaussian white noise can be
made rigorous if the chaotic force differences
act on a relatively small time
scale as compared to $\gamma^{-1}$ and if they have strong mixing
properties.

Next, one knows that in turbulent flows the energy dissipation
$\epsilon$ fluctuates in space and time. In our simple model, the
dissipation process is described by the damping
constant $\gamma$. It is thus most naturally to assume that the
parameter $\beta$ defined as $\beta :=\gamma /\sigma^2$ is a
simple function of the fluctuating energy dissipation in the flow.
For a while, there is local relaxation (energy dissipation) with a
certain value of $\beta =\gamma /\sigma^2 $, then this parameter
changes to a
new value, and so on. 

So unlike an ordinary Brownian particle, in the turbulence
application $u$ is not velocity but velocity difference, moreover
$\beta$ is not inverse temperature but a function of the
fluctuating energy dissipation in the flow. Finally, the correct
interpretation of $E$ is that of an effective potential generating
the relaxation dynamics of $u$, so for example
$E=V(u)=\frac{1}{2}u^2$ generates a linear relaxation dynamics,
whereas other functions $V(u)$ generate more complicated
relaxation processes.

All one has to decide now is what the probability density of the
parameter $\beta$ should be. It is known since the early papers by
Kolmogorov in 1962 that it is reasonable to assume that the
probability density of energy dissipation is approximately
log-normal in a turbulent flow. Hence, if $\beta$ is a simple
power-law function of $\epsilon$, this implies a lognormally
distributed $\beta$. We thus end up in a most natural way with
log-normal superstatistics. If $\beta$ is a more complicated function
of $\epsilon$,
we end up with other superstatistics.

The aim of our simple superstatistics models is neither to solve the
turbulence problem nor to fully reproduce the spatio-temporal
dynamics of the Navier-Stokes equation, but to have a very simple
model that grasps some of the most important statistical
properties of turbulence and at the same time is analytically
tractable.

\subsection{$\chi^2$-superstatistics}

In \cite{prl} a $\chi^2$-distribution was chosen for $f(\beta)$,
\begin{equation}
f (\beta) = \frac{1}{\Gamma \left( \frac{n}{2} \right)} \left\{
\frac{n}{2\beta_0}\right\}^{\frac{n}{2}} \beta^{\frac{n}{2}-1}
\exp\left\{-\frac{n\beta}{2\beta_0} \right\} \label{fluc}
\end{equation}
Here $\beta_0=\int_0^\infty\beta f(\beta ) d\beta$ is the average
value of the fluctuating $\beta$, and $n$ is a parameter
of the $\chi^2$-distribution. For $F(u)=-u$, i.e.\
linear damping forces described by $V(u)=\frac{1}{2}u^2$, the
integral (\ref{9}) is easily evaluated, and one obtains the result
that the marginal distribution $p(u)$ is given by a Tsallis
distribution
\begin{equation}
p(u) \sim \frac{1}{\left(
1+\frac{1}{2}\tilde{\beta}(q-1)u^2\right)^{\frac{1}{q-1}}},
\end{equation}
where the relation between the Tsallis parameters $q$,
$\tilde{\beta}$ and the parameters $n$, $\beta_0$ of the
$\chi^2$-distribution is
\begin{eqnarray}
q&=&1+\frac{2}{n+1}\label{qnn}  \\
\tilde{\beta}&=&\frac{2}{3-q}
\beta_0.
\end{eqnarray}
The distribution has variance 1 for the choice
$\tilde{\beta}=2/(5-3q)$.

In turbulent flows, the assumption of a simple linear damping
force may not be justified. More complicated nonlinear drift
forces may effectively act. If these forces are effectively
described by power-law potentials of the form $V(u)\sim
|u|^{2\alpha}$ one obtains for the marginal density $p(u)$ Tsallis
distributions of the form
\begin{equation}
p(u)=\frac{1}{Z_q}\frac{1}{(1+(q-1)\tilde{\beta}
|u|^{2\alpha})^{\frac{1}{q-1}}} .\label{pu}
\end{equation}
Formulas of this type were shown to very well fit densities
of velocity differences $u$ measured in a Taylor-Couette experiment
\cite{BLS}. Empirically one observes that the relation $\alpha =2-q$
is satisfied by the experimentally measured densities in this experiment.
Using this relation, only one fitting
parameter $q$ remains, which is a function of the scale $r$ on
which the velocity differences are measured and of the Reynolds
number. Excellent fits were obtained for all spatial scales and all
accessible Reynolds numbers. The slight asymmetry of the measured
distributions
can be understood as well \cite{hydro, hilgers1, BLS}. 

\subsection{Log-normal superstatistics}

Let us now proceed to log-normally distributed $\beta$. The
log-normal distribution is given by
\begin{equation}
f(\beta) = \frac{1}{\beta s \sqrt{2\pi}}\exp\left\{ \frac{-(\log
\frac{\beta}{\mu})^2}{2s^2}\right\},
\end{equation}
where $\mu$ and $s$ are parameters. The average $\beta_0$ of the
above log-normal distribution is given by $\beta_0=\mu\sqrt{w}$
and the variance by $\sigma^2=\mu^2w(w-1)$, where $w:= e^{s^2}$.
Let us for the moment restrict ourselves to linear forces
$F(u)=-u$. The integral given by (\ref{9})
\begin{equation}
p(u) = \frac{1}{2\pi s }\int_0^\infty d\beta \; \beta^{-1/2}
\exp\left\{ \frac{-(\log \frac{\beta}{\mu})^2}{2s^2}\right\}
e^{-\frac{1}{2}\beta u^2}  \label{10}
\end{equation}
is the theoretical prediction for the stationary distribution of
velocity differences in the turbulent flow if log-normal
superstatistics is the correct model. The integral cannot be
evaluated in closed form, but the equation is easily numerically
integrated, and can be compared with experimentally measured
densities $p(u)$. The distribution $p(u)$ has variance 1 for the
choice $\mu=\sqrt{w}$, hence only one parameter $s^2$ remains if one
compares with experimental data sets that have variance 1.

The moments for the log-normal superstatistics distribution
(\ref{10}) can be easily evaluated. All moments exist. The
moments of a Gaussian distribution of variance $\beta^{-1}$ are
given by
\begin{equation}
\langle u^m \rangle_G=\frac{1}{\beta^{m/2}} (m-1)!! \label{gauss}
\end{equation} ($m$ even). Moreover, the moments of the lognormal
distribution are given by
\begin{equation}
\langle \beta^m \rangle_{LN} = \mu^m w^{\frac{1}{2}m^2}.
\label{logno}
\end{equation}
Combining eq.(\ref{gauss}) and (\ref{logno}) one obtains the
moments of the superstatistics distribution $p(u)$ as
\begin{eqnarray}
\langle u^m \rangle &=&\langle \langle u^m\rangle_G\rangle_{LN}
\\ &=& (m-1)!! \langle \beta^{-m/2}\rangle_{LN}
\\
&=&(m-1)!! \mu^{-\frac{m}{2}} w^{\frac{1}{8}m^2} \label{r2}
\end{eqnarray}
The variance is given by
\begin{equation}
\langle u^2 \rangle = \mu^{-1} \sqrt{w}.
\end{equation}
All hyperflatness factors $F_m$ are independendent of $\mu$ and
given by
\begin{equation}
F_m:=\frac{\langle u^{2m} \rangle}{\langle u^2 \rangle^m}=
(2m-1)!! w^{\frac{1}{2}(m-1)}.
\end{equation}
In particular, the flatness $F_2$ is given by
\begin{equation}
F_2:=\frac{\langle u^4\rangle}{\langle u^2\rangle}=3w=3e^{s^2}. \label{flat}
\end{equation}
Measuring the flatness $F_2$ of some experimental data thus
provides a very simple method to determine the fitting parameter
$s^2$ of lognormal superstatistics. 

In some recent work \cite{reynolds, my-own}, log-normal
superstatistics and the generalized Langevin dynamics (\ref{1}) is
related  to a generalized Sawford model 
for Lagrangian accelerations 
\cite{saw,pope}. This yields a power-law relation
between $\epsilon$ and $\beta$. The relevance of
distributions of similar form as in eq.~(\ref{10}) has also been emphasized
in early work of Castaing et al. \cite{cast}.

\subsection{Other superstatistics}

In principle, all kinds of distributions $f(\beta)$ can be
considered, leading to different superstatistics models. Which
distribution $f(\beta)$ is the most suitable one, depends on the
physical problem under consideration. As mentioned above, for
turbulent flows there are some arguments that $f(\beta)$ should be
approximately log-normal. The log-normal distribution is probably
still an approximation, it is not the last word, so presumably
there are again some deviations from this and the ultimate
superstatistics model that is the most relevant one to describe
high-Reynolds number 3-dimensional turbulence is simply not known yet.
Nevertheless, for any superstatistics one can define generalized
entropies and (at least in principle) proceed to a generalized
statistical mechanics description, 
following the ideas of \cite{souza}. A turbulent
flow, by construction, is then a complex system of generalized
statistical mechanics that maximizes the above generalized
entropies subject to suitable constraints.

An interesting point is that all superstatistics reduce to Tsallis
statistics for small effective energies $E$: For small $E$
they all have the same quadratic first-order correction to the
ordinary Boltzmann factor. This can be easily seen as follows. For
any distribution $f(\beta)$ with average $\beta_0 :=\langle \beta
\rangle$ and variance $\sigma^2:=\langle \beta^2 \rangle
-\beta_0^2$ we can write
\begin{eqnarray}
B&=& \langle e^{-\beta E} \rangle \\ &=& e^{-\beta_0 E}
e^{+\beta_0 E} \langle e^{-\beta E}\rangle   \nonumber
\\ &=& e^{-\beta_0 E} \langle e^{-(\beta -\beta_0)E}\rangle
\nonumber \\ &=& e^{-\beta_0 E} \left(1 +\frac{1}{2}\sigma^2 E^2
+\sum_{m=3}^\infty \frac{(-1)^m}{m!} \langle (\beta-\beta_0)^m
\rangle E^m \right).
\end{eqnarray}
Here the coefficients of the powers $E^m$ are the $m$-th moments
of the distribution $f(\beta)$ about the mean, which can be
expressed in terms of the ordinary moments as
\begin{equation}
\langle (\beta-\beta_0)^m \rangle =\sum_{j=0}^m \left(
\begin{array}{c} m \\ j \end{array} \right)
\langle \beta^j \rangle (-\beta_0)^{m-j} .
\end{equation}
We see that for small $E$ all superstatistics have a quadratic
correction term to the ordinary Boltzmann factor, and the
coefficient is the same as for Tsallis statistics ($=$
$\chi^2$-superstatistics) if the distribution $f(\beta)$ is chosen
with the same variance $\sigma^2$. In practice, one observes this
'universality' even for moderately large $E$: Many superstatistics
are observed to yield pretty similar results $p(E)$ for moderately
large $E$ (see next section). Usually one observes significant
differences only for very large values of $E$.

\section{Comparison with experiments}

\subsection{Swinney's
data on Taylor-Couette flow}

Figs.~1 and 2 shows an experimentally measured $p(u)$ of velocity
differences $u$ at scale $r=92.5 \eta$ in a Taylor-Couette flow
\cite{BLS}.
$\eta$ denotes the Kolmogorov length scale. The data have been
rescaled to variance 1. The Taylor scale Reynolds number is
$R_\lambda =262$. Apparently, there is excellent agreement between
the measured density and log-normal superstatistics 
as given by eq.~(\ref{10}). The fitting parameter for this
example is $s^2=0.28$.
Note that $s^2$
is the only fitting
parameter. The scale- and Reynolds number
dependence of $s^2$ can be easily extracted from the
measured flatness of the distributions, using eq.~(\ref{flat}).

Fig.~3 shows that essentially the
same curve as for log-normal superstatistics
can be also obtained if one uses Tsallis statistics, i.e.\
eq.~(\ref{pu}) with $q=1.11$ and $\alpha =2-q$. Indeed, the two
theoretical curves can hardly be distinguished in the
experimentally relevant region of $|u|<8$. Significant differences
only arise for much larger $|u|$. So both types of superstatistics
are compatible with the experimental data.

One theoretical advantage of log-normal superstatistics is that it
does not require a nonlinear force $F(u)$, i.e.\ an $\alpha$
different from 1, to fit the data perfectly. A linear forcing is
completely sufficient in that case. The only fitting parameter
that we use for the log-normal superstatistics is $s^2$, since the
parameter $\mu$ is fixed as $\mu = e^{\frac{1}{2}s^2}$ to give
variance 1.

Swinney et al.\ have also measured the probability distribution
of the shear stress $S$ at the outer and inner cylinder in their
Taylor-Couette experiment \cite{swinney}. This distribution is well approximated
by a log-normal distribution, at least for large values
of $S$ (see Fig.~4). The square of the shear stress is essentially the energy
dissipation $\epsilon$ in the flow, and if
\begin{equation}
\beta = C \cdot \epsilon^\kappa \label{pl}
\end{equation}
is some simple power-law function of $\epsilon$ then the measurements of
the nearly log-normal shear stress distribution indicate that the
superstatistics parameter $\beta$ should be approximately
log-normally distributed as well. There are indeed some
theoretical arguments that suggest a power-law relation of type
(\ref{pl}), e.g.\ with $\kappa =-3/2$,
see \cite{my-own} for details. Hence Swinney's measurements
are an indirect experimental hint towards the physical relevance of
log-normal superstatistics.

\subsection{Bodenschatz's data on Lagrangian accelerations}

Accelerations $a$ of Lagrangian test particles in turbulent flows are in
practice measured as velocity differences $u$ on a small time
scale $\tau$ of the order $\tau_\eta$, the Kolmogorov time. Hence
$a\approx u/ \tau$. Fig.~5 shows the most recent measurements of
histograms of accelerations of Lagrangian test particles as
obtained in the experiment of Bodenschatz et al. 
\cite{boden1,boden2,boden3}. The Reynolds number is
$R_\lambda=690$. The
measured distributions are reasonably well approximated by Tsallis
distributions of type (\ref{pu}) for moderately large accelerations (see
e.g. \cite{pla} for a comparison with $\alpha=1$ and Fig.~5 for a comparison
with $\alpha=0.5$).
But for extremely large
accelerations the data seem to systematically fall below curves
corresponding to Tsallis statistics, at least if
the exponent $\alpha$ of the potential $V$
is kept in the physically reasonable range 
$\frac{1}{2}\leq \alpha \leq 1$. As shown in Fig.~5 as well,
log-normal superstatistics provides a better fit of the
tails, with $s^2=3.0$ and using just a linear damping force,
i.e.\ $\alpha =1$. Since Bodenschatz's data reach rather
large accelerations $a$ (in units of the standard deviation), the
measured tails of the distributions allow for a sensitive distinction
between various superstatistics. The main difference between
$\chi^2$-superstatistics and log-normal superstatistics is the
fact that $p(a)$ decays with a power law for the former ones,
whereas it decays with a more complicated logarithmic law for
the latter ones. For alternative fitting approaches, see \cite{ari2}.

One remark is at order. The acceleration is actually
experimentally determined as a parabolic fit of the measured
position of the test particle on a finite time scale $\tau$ of the
order of $\tau_\eta$ , or as a velocity difference on the same
time scale. While in the early paper of the Bodenschatz group
\cite{boden1} no dependence of the data on $\tau$ was mentioned, in the
later paper \cite{boden2} a significant dependence of the flatness of
the distributions on $\tau$ was described (Fig.~28 in \cite{boden2}). The
flatness of the distribution is significantly linked to the tails,
larger flatness certainly means tails that lie higher. So the
shape of all measured distributions actually depends on the
seemingly arbitrary parameter $\tau$. What the asymptotics is for
$\tau \to 0$ depends on extrapolation assumptions. Even the
existence of this limit is not clear at all. In addition, the
tails will presumably still change shape with increasing Reynolds
number. We are theoretically interested in the infinite Reynolds
number case, which could still be very different from the
finite-Reynolds number case. The infinite Reynolds number case
could still be correctly described by Tsallis statistics. And
finally, does a finite-size test particle in the experimental flow
really follow completely the extremely strong forces in the flow,
which are supposed to accelerate it to accelerations of up to
2000 $g$? No astronaut would survive this! So it may well be 
that the measured extreme tails of $p(a)$ contain some
systematic negative corrections, simply because the particle cannot
follow those extreme forces. Summa summarum, one should be very
cautious when drawing over-ambitious conclusions that are solely
based on fits of extreme tail data. The tails describe
acceleration events that are a million times more unlikely than
events near the maximum of the distribution.

\subsection{Gotoh's DNS data} 

Fig.~6 shows Gotoh's results on the
pressure distribution as obtained by direct numerical simulation
of the Navier-Stokes equation at $R_\lambda=380$ \cite{gotoh}. A direct
numerical simulation is also a kind of experiment, just that it is
done on a computer. One usually assumes that in reasonably good
approximation the pressure statistics coincides with the
acceleration statistics of a Lagrangian test particle. Gotoh's
histograms reach accelerations up to 150 (in units of the standard
deviation), a much larger statistics than can be presently reached
in Bodenschatz's experiment. Hence the tails of these
distributions can very sensitively be used to distinguish various
superstatistics models. 

Fig.~6 shows that log-normal
superstatistics with $s^2=3.0$ and linear forcing
again yields a good fit of the tails,
keeping in mind that one compares data that vary over 12 orders of
magnitude. 

Near the maximum of the distributions, the fit quality of log-normal
superstatistics 
is not very good: $p(0)$ is too big as compared to the DNS data.
But this defect can be easily 
cured by introducing an upper cutoff in $\beta$. That is to say,
in eq.~(\ref{10}) we only integrate up to a certain $\beta_{max}$ and
re-normalize afterwards.
Log-normal
superstatistics
with an upper cutoff of $\beta_{max}\approx 32$ yields quite a perfect
fit in the vicinity of the maximum (Fig.~7). The tails
are not influenced by this cutoff. 
The above truncation may
effectively represent finite size or finite Reynolds number
effects, 
which are certainly present in any numerical simulation of the
Navier-Stokes equation.

As Fig.~8 shows, Tsallis statistics
with $q=1.476$ and $\alpha =0.832$
also yields a very good fit of the
data in the vicinity of the maximum (the relation between $q$ and
$\alpha$ is $q=1+2\alpha/(3\alpha +1)$, the theoretical prediction of the model
considered in \cite{prl} with $n=3$
spatial dimensions). But for very large accelerations
Tsallis statistics implies power-law tails, which are not
supported by the finite-Reynolds number DNS data. 

Of course the following general question arises:
How much can we believe in the extreme tails of a DNS simulation?
It should be clear that every DNS is a brute force finite
lattice size approximation of the Navier-Stokes equation.
Naturally there are finite-lattice size effects, also finite
lattice constant effects, and moreover
finite Reynolds number effects, which
may heavily influence the extreme tails. Moreover, do the extreme
events of 150 standard deviations, corresponding to accelerations
of almost $10000\; g$, really describe plausible physics? Can a true
physical test particle really follow such a force? Bodenschatz's
experiment, tracking single test particles, leads to $p(a)\sim
10^{-8}$ for the most rare acceleration events. The detector for
these measurements was running for about a month to collect the
data. Gotoh's DNS data reach $p(a)\sim 10^{-12}$ for the rarest
acceleration events. This statistics is larger by a factor $10^4$.
Hence Bodenschatz, in a laboratory experiment similar to the one
he did so far, would need to wait $10^4$ months $\approx 1000$
years to observe one of the extreme acceleration events described
by Gotoh's numerical simulation. I guess most physicists are not
willing to wait that long.

\section{Conclusion}

By analyzing three different data sets obtained by Swinney,
Bodenschatz, Gotoh, respectively, we have shown that measured and
simulated  densities in Eulerian and Lagrangian turbulence 
experiments are
well described by simple superstatistics models. Log-normal
superstatistics differs from $\chi^2$-superstatistics, i.e.\
ordinary Tsallis statistics, but for moderately large velocity
differences log-normal superstatistics
can be quite close to Tsallis
statistics, as shown e.g.\ in
Fig.~3. The fact that this is so is not surprising but simply
a consequence of the `universality' property discussed in section
2.5. For small effective energies $E$ (i.e. small $u$ or $a$ in
the turbulence application) {\em any} superstatistics is close to
Tsallis statistics. In practice we see that this is often also the
case for moderately large velocity differences and accelerations.
Significant differences only arise for very large velocity
differences (and large accelerations), where Tsallis statistics
predicts a power law decay of probability density functions,
whereas log-normal superstatistics yields tails that decay in a
more complicated way. It is indeed the tails contain the
information on the most appropriate superstatistics for turbulent
flows. A precise estimate of the error bars of the tails of
experimentally measured or simulated
distributions is clearly needed, taking into account not
only statistical errors but all systematic errors as well.
Moreover, one would wish for precise data on
how the shape of the tails depends on the time scale
on which the accelerations are measured, and how the tails
change
with Reynolds number. Finally, it would be interesting to
have precise data on correlation functions of accelerations,
since these yield more information than the densities alone.

\subsection*{Acknowledgement}
I am very grateful to Harry Swinney, Eberhard Bodenschatz and
Toshiyuki Gotoh for providing me with the experimental data
displayed in the various figures.

\newpage

\epsfig{file=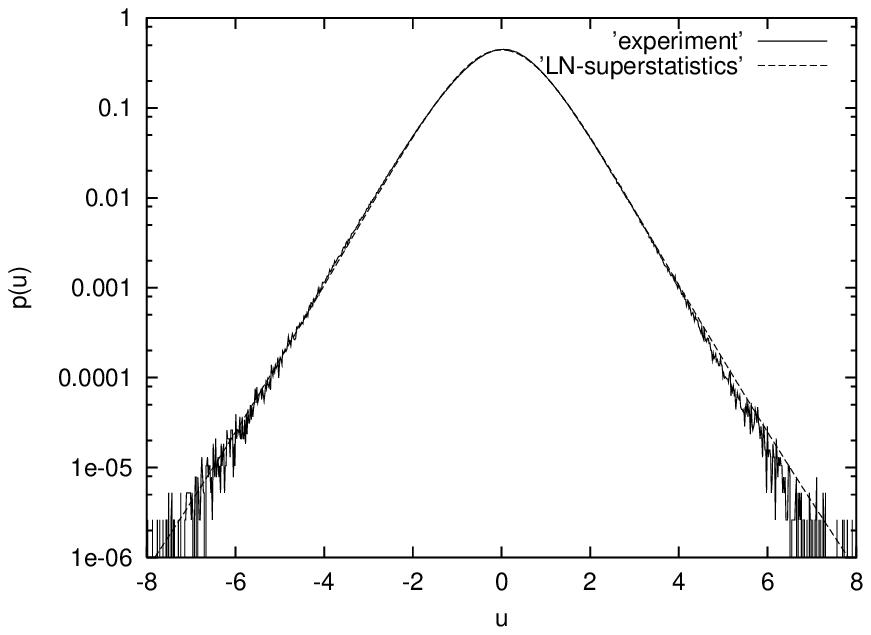}

{\bf Fig.~1} Histogram of velocity differences $u$ as measured
in Swinney's experiment and the 
log-normal superstatistics prediction eq.~(\ref{10}) with $s^2=0.28$.

\vspace{1cm}

\epsfig{file=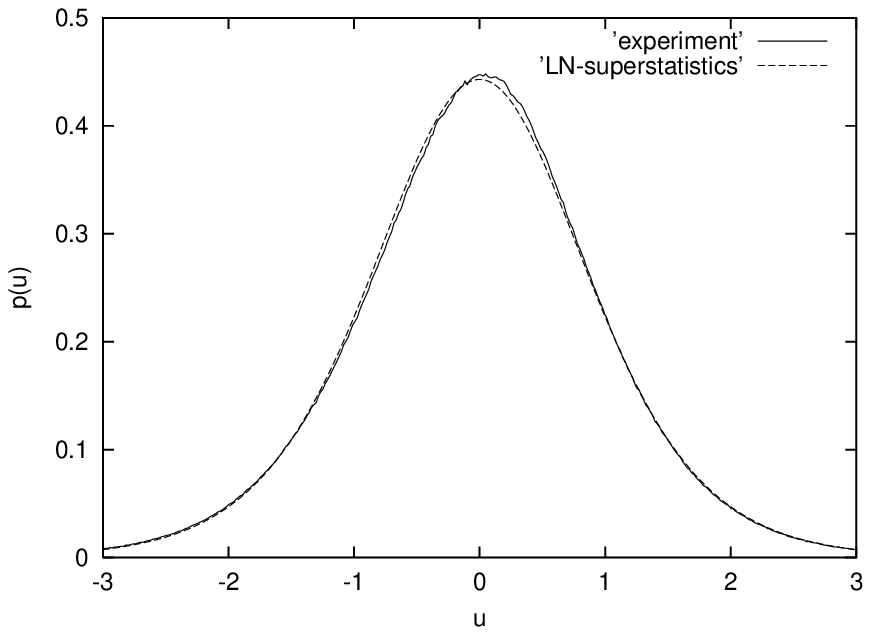}

{\bf Fig.~2} Same as Fig.~1, but a linear scale is chosen. This
emphasizes the vicinity of the maximum, rather than the tails.

\vspace{1cm}

\epsfig{file=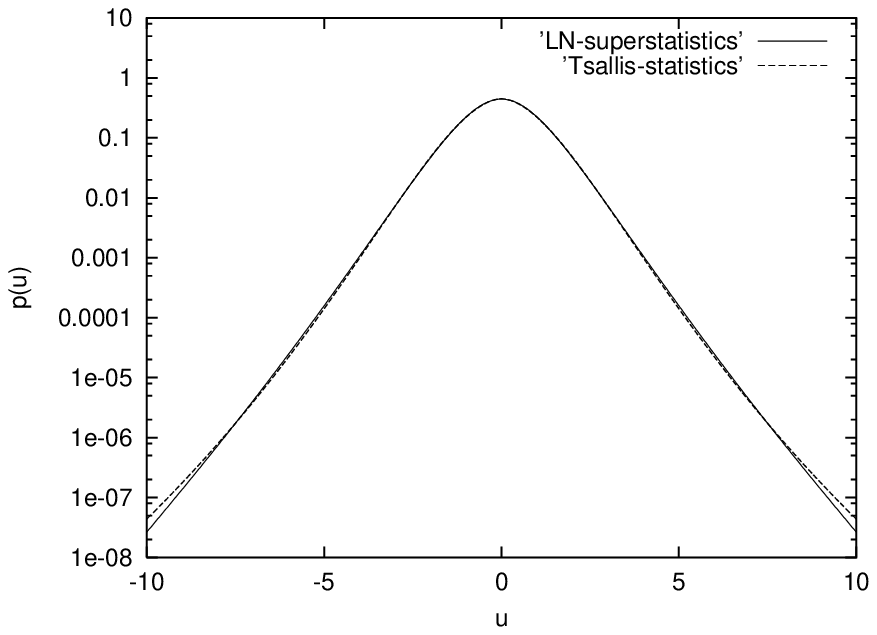}

{\bf Fig.~3} Comparison between log-normal superstatistics as
given by eq.~(\ref{10})
with $s^2=0.28$ and Tsallis statistics as given by
eq.~(\ref{pu}) with $q=1.11$
and $\alpha=2-q$. For the range of values accessible in the experiment,
$|u|<8$,
there is no visible difference between the two curves.

\vspace{1cm}

\epsfig{file=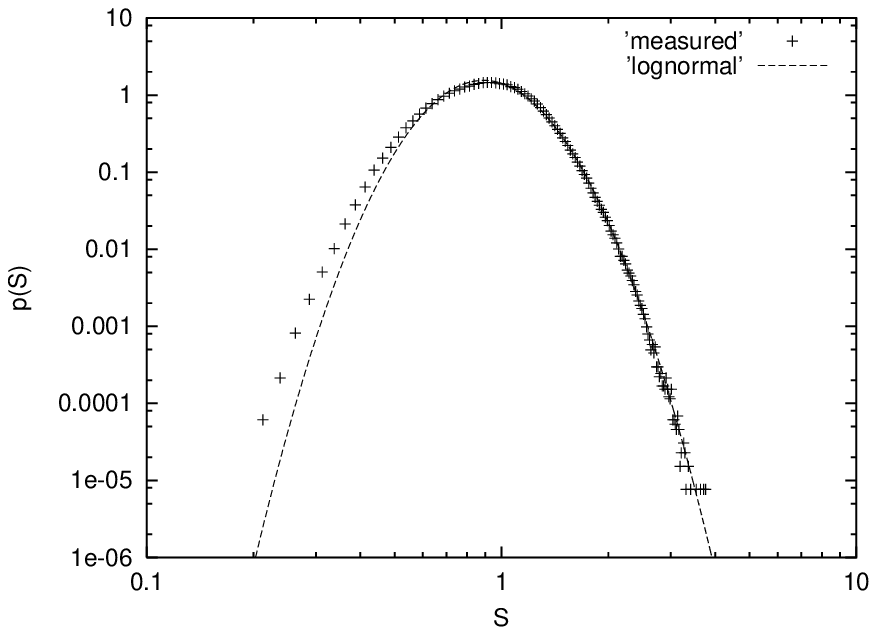}

{\bf Fig.~4} Swinney's measurements
of the shear stress distribution at the outer cylinder
of the Taylor-Couette experiment, and 
comparison with a log-normal distribution.

\vspace{1cm}

\epsfig{file=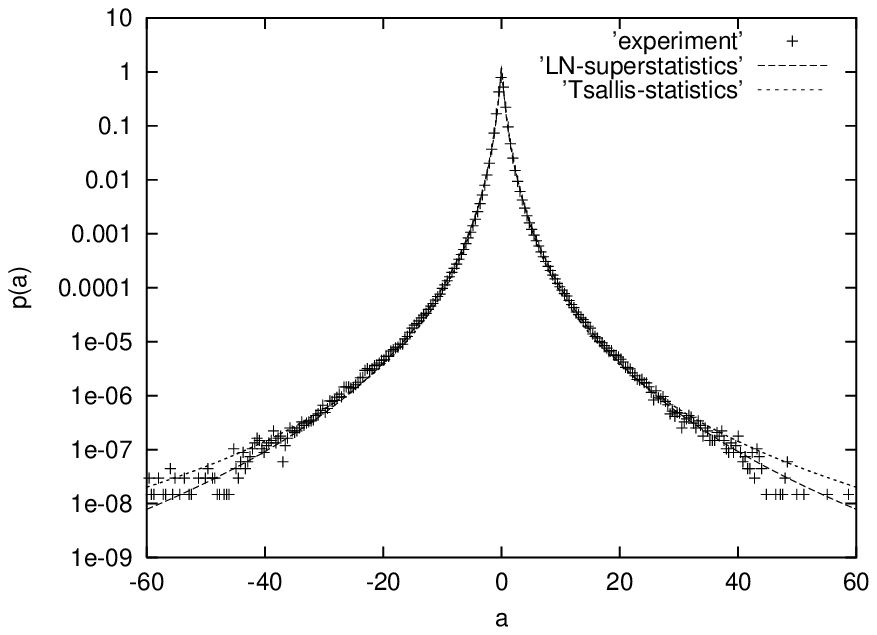}

{\bf Fig.~5} Acceleration distribution as measured by Bodenschatz et al.\
and comparison with the log-normal superstatistics distribution (\ref{10})
with $s^2=3.0$. Also shown is a Tsallis distribution (\ref{pu})
with $q=1.2$ and
$\alpha=0.5$.

\vspace{1cm}

\epsfig{file=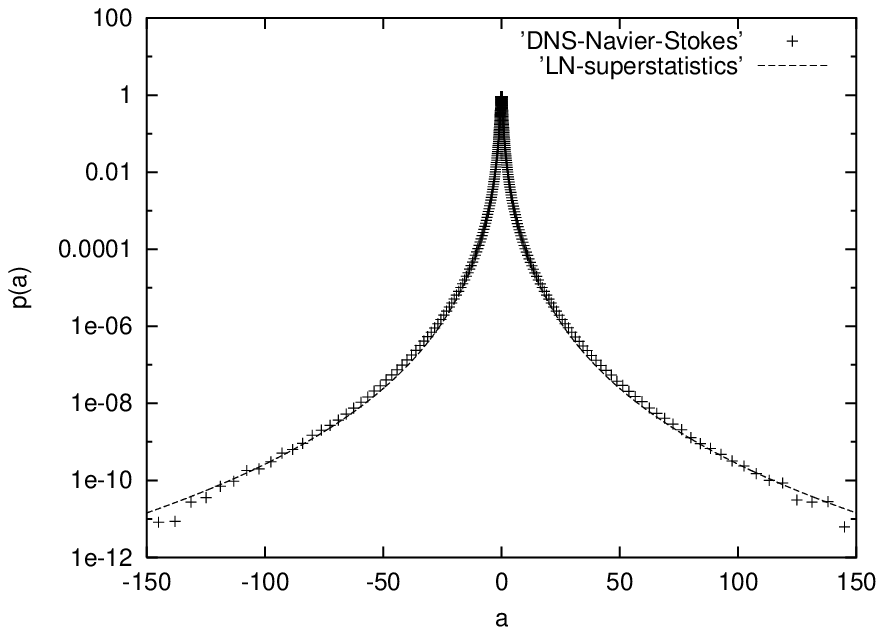}

{\bf Fig.~6} Pressure statistics as obtained by Gotoh et al.
in a direct numerical simulation of the Navier-Stokes equation, and comparison
with log-normal superstatistics with $s^2=3.0$.

\vspace{1cm}

\epsfig{file=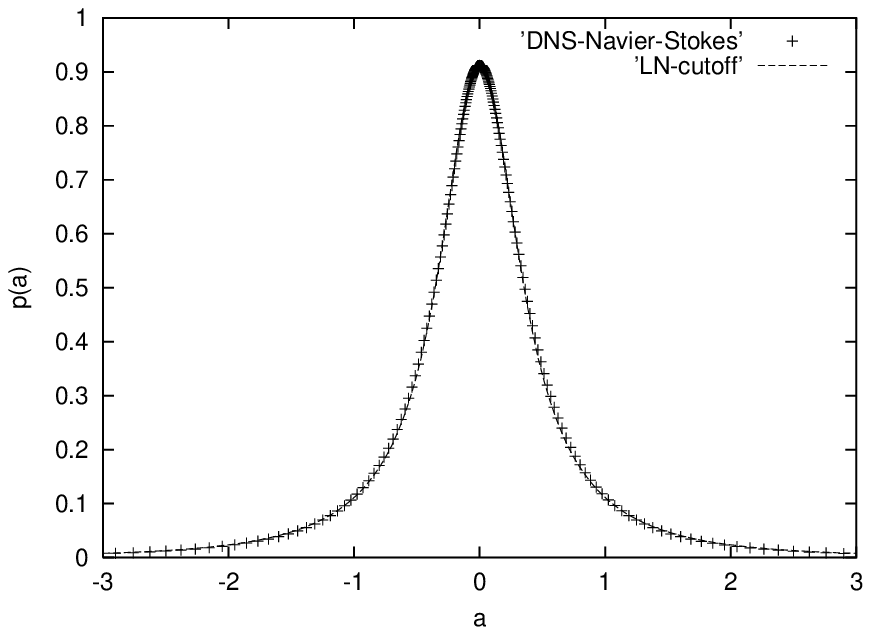}

{\bf Fig.~7} Same data as in Fig.~6, but a linear scale is chosen
to emphasize the vicinity of the maximum. The
fitted line (hardly visible behind the
data points) corresponds to log-normal superstatistics with $s^2=3.0$
and an upper cutoff $\beta_{max}=32$.

\vspace{1cm}

\epsfig{file=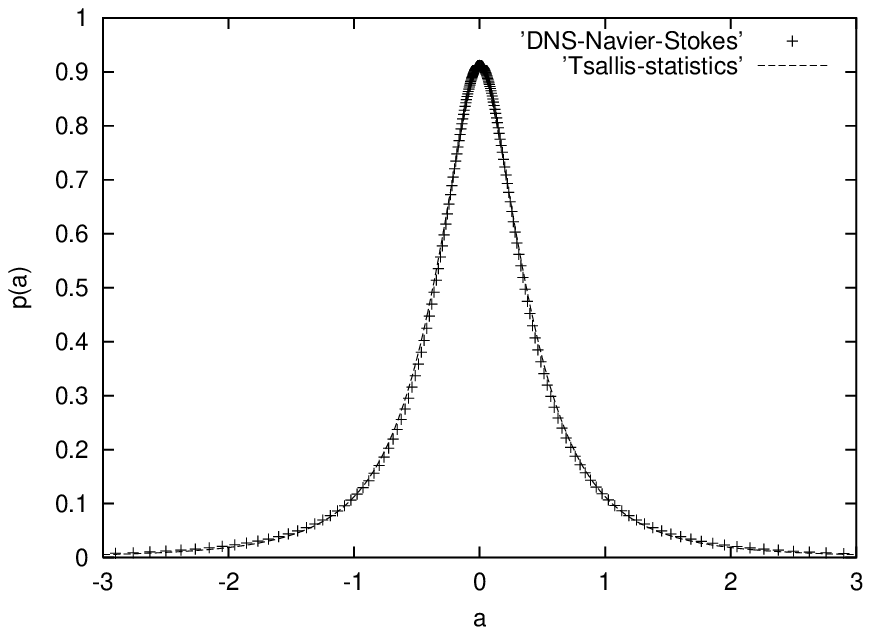}

{\bf Fig.~8} Same data as in Fig.~7.
The fitted line corresponds to Tsallis statistics with $q=1.476$ and
$\alpha=0.832$. Only $\alpha$ is fitted--the value of $q$ follows from 
formula (17) in \cite{prl}.
\end{document}